# On some numerical methods in application to low-Reynolds-number turbulence models

*S.V.Utyuzhnikov*

Department MAME, UMIST, PO Box 88, Manchester, M60 1QD, UK

Department of Computational Mathematics,
Moscow Institute of Physics & Technology, Dolgoprudny 141700, Russia

s.utyuzhnikov@umist.ac.uk

**Abstract**

To study and develop wall-functions for low-Reynolds-number models, a model linear equation is introduced. This equation simulates major mathematical peculiarities of the low-Reynolds-number model including a near wall sub-layer and transition region. Dirichlet and Newman boundary-value problems are considered. The standard and analytical wall-functions are investigated on different properties including the mesh sensitivity of a solution. A Robin-type interpretation of wall functions as boundary conditions is suggested. It is shown that solution of a problem is mesh independent and more accurate in this case. General type analytical and numerical wall-functions are developed on the basis of a boundary condition transfer. An effective numerical method of decomposition is suggested. The method can be used in application to either high-Reynolds-number models with the numerical wall-functions or low-Reynolds-number models directly. Although a model equation is considered, the formulas, methods and conclusions are valid and can be directly used for real low-Reynolds-number equations.

## 1. Introduction

The problem of mathematical simulation of turbulent flows near walls appears in many practical applications. It is well known that turbulence vanishes near a wall due to the no-slip boundary condition for the velocity as well as the blocking effect caused by the wall. In the vicinity of the wall, there is a thin sub-layer with predominantly molecular diffusion. The sub-layer has a substantial influence upon the remaining part of the flow. An adequate numerical resolution of a solution in the sub-layer requires a very fine mesh because of sub-layer thinness and high gradients of the solution. It makes a model used time consuming and often not suitable for industrial applications. Because of low velocities, the models resolving the sub-layer are called the low-Reynolds-number models (LR models).

To avoid the problem related to the sub-layer resolution so-called high-Reynolds-number models (HR models) have been developed. In this a type of models the sub-layer is not directly resolved. It allows one to save computational efforts many times over [1]. In the HR models, the boundary conditions or near-wall profiles are replaced by wall-functions. In most cases, the wall-functions are semi-empirical and have very limited applications [1-3]. A brief review of different wall-functions used can be found in, e.g. [1]. Sub-grid numerical wall-functions are developed in [2] where the dependent variables are determined by solving boundary-layer-type transport equations in a sub-grid. In this approach, the boundary condition on the boundary that is external to the wall is determined by linear interpolation of certain main-grid values. In [3], the analytical wall-functions are evolved by integrating boundary-layer-type equations analytically under some simplifying assumptions. At the wall the boundary conditions are the same as the ones used in the LR models. Then, the analytical profiles are used in the cell nearest to the wall to reconstruct the near-wall solution.

In the following sections we study the analytical wall-function approach [3] for the case of a model linear equation. This model equation allows us to simulate the major mathematical peculiarities of LR models. A method of boundary condition transfer is developed. The method allows us to move the boundary conditions from the wall outside of the sub-layer. The boundary conditions developed are Robin type and can be interpreted as wall-functions. It is possible to obtain such boundary conditions both analytically and numerically. In the former case the boundary conditions can be obtained exactly. A decomposition method is also suggested. The method allows us to split the boundary-value problem into two problems: a problem internal to the wall and an external one. Both boundary-value problems are solved independently, which yields the terminal solution.

## 2. Model equation

Considering the following model equation:

$$(\mu u_y)_y + y^n u_y = C, \qquad (1)$$

defined in a domain $\Omega = [0\ 1]$.

Where $\mu = (1 - exp(-y/\varepsilon) + \delta)/Re$, $\varepsilon << 1$, $\delta << 1$, $Re >> 1$, $n > 0$.

The first term simulates dissipative terms in the Navier-Stokes equations, whereas the second term models the contribution of convective terms and the right hand side represents the pressure gradient term or source in the transport equations. The "viscosity" coefficient $\mu$ corresponds to the effective viscosity coefficient. The coefficient is rapidly changed from a relatively small value $\mu_l = \delta/Re$ (laminar viscosity) to a "turbulent" viscosity $\mu_t \approx 1/Re$.

The equation simulates the low-Reynolds-number effects and can be considered as a model equation for the LR model. The left hand side point in the domain $\Omega$ will be treated as a "wall". The low-Reynolds-number effects occur nearby this point. If we set $\mu = (1+\delta)/Re$ in (1), we have the HR approach.

Furthermore, we will consider the following values for the constants: $Re = 10^2$, $\varepsilon = 3*10^{-2}$, $\delta = 10^{-2}$. For simplicity the right-hand side will be considered as the constant $C = -1$ although this assumption is not important for our consideration.

Consider the following boundary-value Dirichlet problem:

$$(\mu u_y)_y + y^n u_y = C \qquad (2)$$
$$u(0) = u_0 \quad u(1) = u_1$$

In the case of $n = 2$, $u_0 = 0$ and $u_1 = 7$, the solution is shown in Figure 1. The profile includes both the linear near-wall and logarithmic parts. Near the wall, $u = \varepsilon \ln(1+y/(\varepsilon\delta))$. The thickness of the viscous "sub-layer", where $u$ is a linear function, can be approximated as $y_v \approx \varepsilon\delta$.

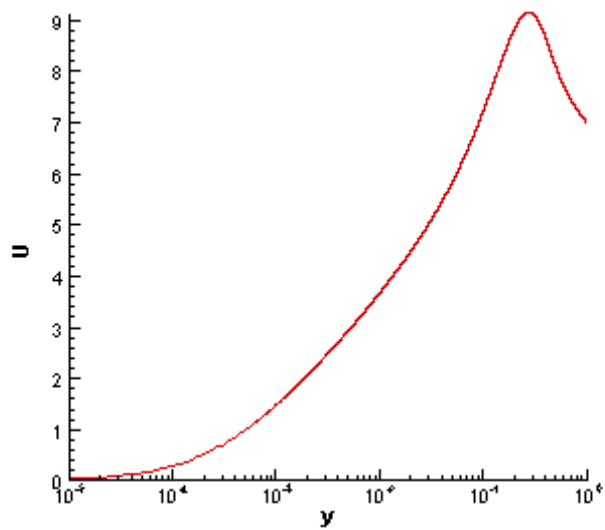

Fig.1 The exact solution.

In the calculations, the compact scheme [4] is used. The scheme allows us to calculate both the function and its derivative with a fourth-order of approximation. In Table 1, the results of calculations of the "friction" $\tau_w = \mu_w du/dy(0)$ (here $\mu_w = \mu(0)$) are given on different meshes. To exclude questions concerned with a mesh adaptation, a uniform mesh is used.

| Grid | $10^2$ | $5*10^2$ | $10^3$ | $5*10^3$ | $10^4$ |
|---|---|---|---|---|---|
| $\tau_w$ | $3.1*10^{-1}$ | $3.2*10^{-1}$ | $3.3*10^{-1}$ | $3.36*10^{-1}$ | $3.36*10^{-1}$ |

Table 1. Calculation of $\tau_w$ on different meshes.

If we use the HR model with the same boundary conditions, then $\tau_w = 4.9*10^{-1}$, which shows the importance of the sub-layer.

## 3. Wall functions

*3.1 Standard wall-function*

To use the HR model, the wall boundary condition can be substituted by wall-functions. In this case, we set the boundary condition outside the sub-layer. In fluid mechanics, the classic wall-function is given by the law of the wall. It corresponds to the log-profile. In our case, it means the following local relation:

$$u = \tau_w \varepsilon \ln(\frac{y}{\varepsilon\delta}) + \varsigma, \qquad (3)$$

where $\varsigma$ is a constant defined from experiments. We assume that it is equal to zero.

*3.2 Analytical wall-function*

In [3], the analytical wall functions have been developed. To obtain them, the governing equation is integrated in the vicinity of a wall under the assumption that all terms besides the dissipative one are constant. Mainly, it means that the contribution of the convective terms is neglected near the wall and that the pressure gradient and buoyancy force (if applicable) are not changed. In this case, the following equation is integrated

$$(\mu u_y)_y = C \qquad (4)$$

Following [3], we assume a linear approximation for the viscosity $\mu$ in the sub-layer:

$$\mu^{-1} = \text{Re}(1 + b_\mu(y - y_v))/(1+\delta), \quad b_\mu = \frac{-1}{\delta y_v} \qquad (5)$$

Outside of the sub-layer $\mu = (1 + \delta)/\text{Re}$.

Such an approximation looks reasonable. The error in the 1$^{st}$ norm is as follows:

$$\| \mu_{ex} - \mu_{ap} \|_1 \approx 10^{-2}$$

The difference between the exact and approximate values of $\mu$ is given in Figure 2 ($y_v = \delta\varepsilon$).

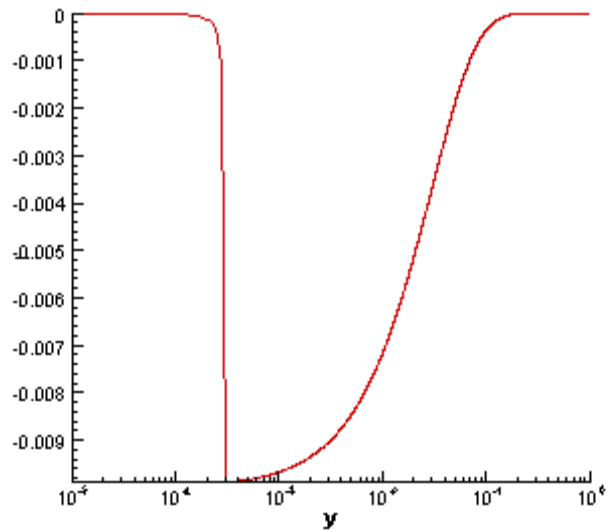

Fig.2. Error function in approximation of μ.

Integrating (4), we obtain:

$$u = \begin{cases} u_0 + \tau_w \text{Re} y \left[1 + b_\mu/2(y - 2y_v)\right] + \dfrac{C\text{Re} y^2}{2}\left[1 - b_\mu y_v + 2/3\, b_\mu y\right] & 0 < y < y_v \\ u_0 + \tau_w \text{Re} y \left[1 - b_\mu/2 * y_v^2/y\right] + \dfrac{C\text{Re} y^2}{2}\left[1 - b_\mu/3 * y_v^3/y^2\right] & y > y_v \end{cases} \quad (6)$$

Here $Re/(1 + \delta) \approx Re$.

In approximate solution (6) there are two integration constants, namely: $u_0$ and $\tau_w$. One of which is known from the boundary condition. Generally speaking, other types of the boundary conditions are possible, e.g., mixed conditions or Robin-type boundary conditions. Only cases of a Dirichlet problem ($u_0$ is known) and a Newman problem ($\tau_w$ is known) are considered because they most common in applications. It is interesting to set the exact values of both constants and compare with the exact solution. Such a comparison is shown in Figure 3 for problem (2). The exact solution (solid line) and the two approximations for $y_v$ are represented: $\delta\varepsilon$ (dotted-dashed line) and $3\delta\varepsilon$ (dashed line). In the latter case ($y_v = 3\delta\varepsilon$) the correspondence with the exact solution is better.

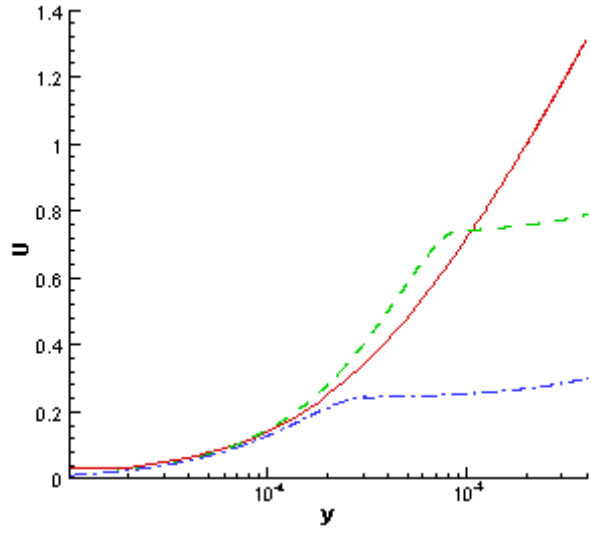

Fig. 3. Near wall profiles. Solid line is exact solution,
dotted-dashed line ($y_v = \delta\varepsilon$), dashed line ($y_v = 3\delta\varepsilon$)

Relations (6) correspond to the analytical wall-functions [3]. According to [3], if $u_0$ is known, the HR model can be used with the boundary condition for $u_0$. The profile in the first computational cell is then completed with (6). In particular, relation (6) allows us to estimate $\tau_w$.

In Table 2, the results of the computation of $\tau_w$ are given on different meshes using the standard wall-function (3) and analytical wall-function (6). In the second row, $y+_1$ is the value of $y+ = \sqrt{\tau_w}\, y/\nu$ ($\nu = \delta/Re$) at the point nearest to the wall. The analytical wall-functions are considered for two values of $y_v$: $y_v = \delta\varepsilon$ (1) and $y_v = 3\delta\varepsilon$ (2).

| Grid | 10 | 20 | 50 | $10^2$ | $10^3$ | $10^4$ |
|---|---|---|---|---|---|---|
| $y+_1$ | $4*10^2$ | $2*10^2$ | 81 | 41 | 4 | $4*10^{-1}$ |
| Standard | $2.5*10^{-1}$ | $1.5*10^{-1}$ | $7.7*10^{-2}$ | $4.6*10^{-2}$ | $1.4*10^{-2}$ | $-2*10^{-3}$ |
| Analytical 1 | $4.2*10^{-1}$ | $3.8*10^{-1}$ | $2.8*10^{-1}$ | $2.0*10^{-1}$ | $3*10^{-2}$ | $6*10^{-3}$ |
| Analytical 2 | $3.3*10^{-1}$ | $2.6*10^{-1}$ | $1.5*10^{-1}$ | $9*10^{-2}$ | $10^{-2}$ | $5*10^{-3}$ |

Table 2. Calculation of $\tau_w$ on different meshes using standard and analytical wall functions. The exact solution is 0.34. Analytical 1 is $y_v = \delta\varepsilon$; analytical 2 is $y_v = 3\delta\varepsilon$.

The following conclusion can be made: the analytical wall-functions provide less mesh dependent solution than the standard wall-functions. Nevertheless, the analytical wall-functions are highly sensitive to the sub-layer thickness $y_v$ and the dependence on mesh is relatively high. In the case of a fine mesh, both wall-function approaches fail and some extra damping terms are necessary [3]. As an alternative, the different interpretation of relation (6) is suggested in the next section.

*3.3. Robin-type treatment of wall functions*

From (6), we have

$$\mu du/dy = \tau_w + C y \tag{7}$$

Excluding $\tau_w$ from (6) by (7), we obtain:

$$u = \begin{cases} u_0 + y\dfrac{du}{dy}\operatorname{Re}\mu\left(1 + b_\mu/2(y - 2y_v)\right) - \dfrac{C\operatorname{Re}y^2}{2}\left(1 + b_\mu(y_1/3 - y_v)\right) & 0 < y < y_v \\ u_0 + y\dfrac{du}{dy}\left(1 - \dfrac{b_\mu y_v^2}{2y}\right) - \dfrac{C\operatorname{Re}y^2}{2}\left(1 - b_\mu y_v^2/y^2(y_1 - y_v/3)\right) & y \geq y_v \end{cases} \tag{8}$$

It is important to emphasize that relations (8) are accurate for under the assumption that one neglects the "convective" term $y^n du/dy$. If we know $u_0$ from the boundary conditions, we can consider equalities (8) as Robin-type boundary conditions for the HR model at any point $y = y^* > 0$. It is reasonable to choose $y^*$ outside the sub-layer. On the other hand, $y^*$ cannot be too far from $0$ since relations (8) are valid only near the wall. Then, we can consider (8) as a boundary condition at the wall.

If we move these equalities to the boundary, we get a Robin-type boundary condition at the wall. This boundary condition is similar to the "slip boundary condition" at the edge of the Knudsen-layer in aerodynamics.

Solving the HR equation with the boundary condition (8), $\tau_w$ satisfies:

$$\tau_w = \frac{\alpha}{\operatorname{Re}}\frac{du}{dy}(0) - Cy^*, \tag{9}$$

where $\alpha = 1 - y^*$ is a scaling coefficient because of moving the boundary condition from point $y^*$ to the wall. It provides some minor correction only if $y^*$ is big enough.

The wall-function (8) does not depend on the mesh used. There is some dependence on $y_v$ and $y^*$ but it is weaker. The calculation results for different values of the parameters $y_v$ and $y^*$ are given in Table 3. It is important that the parameter $y^*$ is not related to the mesh used at all. Therefore, any arbitrary mesh can be used in calculations including the near-wall region.

| $y^*$ | $10^{-1}$ | $5*10^{-2}$ | $10^{-2}$ | $10^{-3}$ | $10^{-4}$ |
|---|---|---|---|---|---|
| $y_v = \delta\varepsilon$ | 0.4 | 0.42 | 0.45 | 0.47 | 0.48 |
| $y_v = 3\delta\varepsilon$ | 0.37 | 0.39 | 0.41 | 0.42 | 0.48 |

Table 3. Calculation of $\tau_w$. Robin-type boundary condition. Exact $\tau_w = 0.34$

The comparison between the exact LR solution, solutions for different values of $y^*$ and the HR solution with the Dirichlet boundary condition (2) is given in Figure 4 ($y_v = 3\,\delta\varepsilon$). If $y^* = 0.1$, outside of the sub-layer the solution almost coincides with the LR solution.

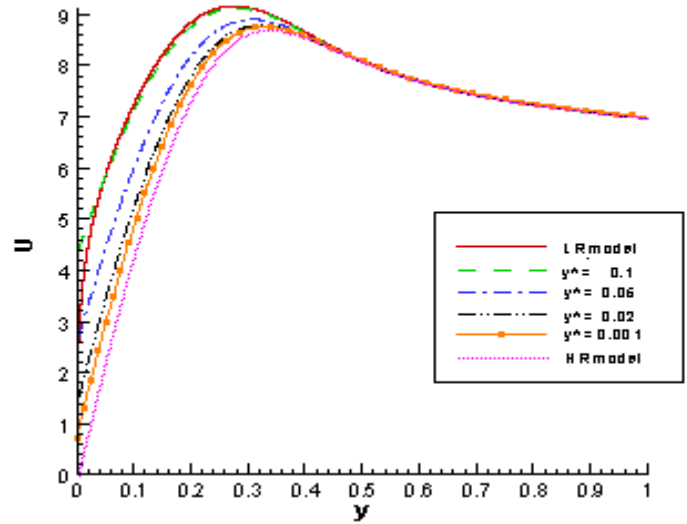

Fig 4. Profiles of $U$ for different $y^*$, LRM solution (solid line) and HRM solution with LRM boundary conditions (dotted line)

It is possible to set the boundary condition at point $y^*$ (or nearest mesh point) in the computational domain. In this case, the boundary-value problem is numerically solved in the domain $y^* < y < 1$. In the rest of the domain, $0 < y < y^*$, the solution can be obtained analytically, and it will be considered in the next section in detail. With this approach, the solution almost does not depend on $y^*$ provided $y^*$ is considered outside the sub-layer. It is very close to the ultimate solution when the boundary condition is determined at the wall and $y^* \approx y_v$.

### 3.4 Generalized wall-functions

In this section, we develop wall-functions in a general case without any approximation of the coefficient $\mu$. It is assumed that the convective term can be neglected in some vicinity of the wall.

After integrating equation (4) from $0$ to $y$, one obtains:

$$u(y) = u_0 + \tau_w \int_0^y \frac{1}{\mu} d\xi + C \int_0^y \frac{\xi}{\mu} d\xi \tag{10}$$

On the other hand, from (7) considering at $y^*$:

$$\tau_w = \mu(y^*) du/dy(y^*) - C y^* \tag{11}$$

Then

$$u(y) = u_0 + \frac{du}{dy}(y^*) \int_0^y \frac{\mu(y^*)}{\mu(\xi)} d\xi - C \int_0^y \frac{(y^* - \xi)}{\mu(\xi)} d\xi \tag{12}$$

Introducing function $\zeta = \mu^*/\mu$ ($\mu^* = \mu(y^*)$), and rewriting (12) as follows:

$$u(y) = u_0 + \frac{du}{dy}(y^*) \int_0^y \zeta \, d\xi - \frac{C}{\mu^*} \int_0^y \zeta (y^* - \xi) \, d\xi \tag{13}$$

If we now introduce $\eta = \dfrac{\mu^*}{\mu} \dfrac{\mu - \mu_w}{\mu^* - \mu_w}$ ($\mu_w = \mu(0)$, $0 \leq \eta \leq 1$),

then

$$\zeta = (1 - \alpha_\mu)\eta + \alpha_\mu, \quad \alpha_\mu = \mu^*/\mu_w$$

Considering (13) at point $y^*$, the following equality is obtained:

$$u(y^*) = u_0 + y^* \frac{du}{dy}(y^*) f_1 - \frac{C y^{*2}}{2\mu^*} f_2, \tag{14}$$

where

$$f_i = \alpha_\mu + (1 - \alpha_\mu) I_i, \quad i = 1, 2$$
$$I_1 = \int_0^1 \eta \, d\xi, \quad I_2 = 2 \int_0^1 \eta (1-\xi) d\xi \quad (\xi = y/y^*) \tag{15}$$

Equality (14) is accurate for any arbitrary integrable function $\mu$ under the assumption that the convective term is negligible. Considering (14) as the Robin type boundary condition at either the wall or point $y^*$ similarly to the previous section. Integrals (15) are estimated either numerically or analytically. The following estimations for $f_i$ are valid: $f_1 > 1, f_2 > 0$.

If $C = C(y)$, then equalities are generalized as follows:

$$u(y^*) = u_0 + y^* \frac{du}{dy}(y^*) f_1 - \frac{y^{*2}}{2\mu^*} f_2 \int_0^1 C d\xi, \qquad (14')$$

where

$$f_i = \alpha_\mu + (1 - \alpha_\mu) I_i, \quad i = 1, 2$$

$$I_1 = \int_0^1 \eta \, d\xi, \quad I_2 = 2 \int_0^1 \eta \left( 1 - \frac{\int_0^\xi C d\xi'}{\int_0^1 C d\xi} \right) d\xi \quad (\xi = y/y^*) \qquad (15')$$

Assuming that the coefficient $\mu$ is changed linearly from $\mu_w$ to $\mu^*$, then ($C = const$):

$$I_1 \approx 1 - \frac{1 + \ln \alpha_\mu}{\alpha_\mu} \quad (\text{if } \alpha_\mu \gg 1), \quad I_2 \approx 1 - \frac{2 \ln \alpha_\mu}{\alpha_\mu}$$

$$f_1 \approx 2 + \ln \alpha_\mu, \quad f_2 \approx 1 + 2 \ln \alpha_\mu$$

If we assume piecewise dependence (5) as in [3], then

if $0 < y < y_v$

$$I_1 = \tfrac{1}{2}, \; I_2 = 1/3, f_1 = (1 + \alpha_\mu)/2, \; f_2 = (1 + 2\alpha_\mu)/3$$

if $y > y_v$

$$I_1 = 1 - \xi_v/2, \; I_2 = 1 - \xi_v + \xi_v^2/3$$

$$f_1 = 1 + (\alpha_\mu - 1)\xi_v/2, \; f_2 = 1 + (\alpha_\mu - 1)(\xi_v - \xi_v^2/3),$$

where $\xi = y_v/y^*$. Relations (8) are shown as a particular case.

In both cases explicit boundary conditions are obtained.

In a general case, if integrals (15) based on the exact value of $\mu$ are estimated numerically, it is possible to develop the boundary condition of a general type for any arbitrary integrable function $\mu$.

As in the previous section, the boundary condition does not depend on a mesh. There is some dependence on $y^*$ only, although it is not too significant. To decrease the dependence on $y^*$, one may assume that

$$\mu^* = \mu_e \approx 1/Re \qquad (16)$$

The results of the computations of $\tau_w$ at different values of $y^*$ are given in Table 4.

| $\tau_w$ | $y^* = 10\varepsilon\,\delta$ | $y^* = 2\varepsilon$ | $y^* = 3\varepsilon$ |
|---|---|---|---|
| $\mu = \mu(y^*)$ | 0.05 | 0.3 | 0.3 |
| $\mu = \mu_e$ | 0.42 | 0.32 | 0.31 |

Table 4. Calculation of $\tau_w$ for different $y^*$ and $\mu^*$. Exact solution is 0.34

Correction (16) is essential only for small enough values. Such values are unrealistic since $y^*$ is to be chosen close to the fully "turbulent" layer ($\mu \approx \mu_e$), if considering the real coefficient $\mu$.

Comparison with the exact solution (solid line) is done in Figure 5 for $y^* = 2\varepsilon$. The dashed and dashed-dotted lines correspond to the boundary condition determined at $y^*$ in $\Omega$; the line with little squares is the version with the boundary condition at the wall. In the former case the solution consists of two parts and very close to the exact solution. The reason for the difference from the exact solution in the latter case is as follows. When the boundary condition is set at the wall, the convective term becomes smaller and its decay influences the solution substantially. The lack of accuracy in the composite solution based on the analytical expressions (7) and (8) (Fig. 4) is explained by the approximation of $\mu$.

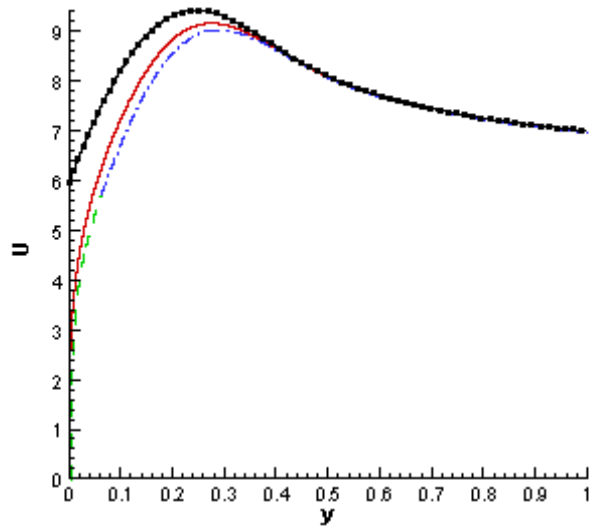

Fig.5. Comparison of the exact (LR) solution and solutions obtained by generalized wall-function.

*3.4.1 Newman problem*

Once the Newman problem is solved, the algorithm is similar, but with the following modifications. The boundary condition at some point $y^*$ is made using (7):

$$\frac{du}{dy}(y^*) = (\tau_w + Cy)/\mu^* \qquad (17)$$

Then, the boundary-value problem is solved either for $0 \le y \le 1$ or $y^* \le y \le 1$. The value of $u(0)$ at the wall for the initial problem can be determined either from (8) or from (14) upon obtaining $u(y^*)$ and $du/dy(y^*)$.

*3.5 Numerical wall-functions and decomposition method*

In this section we develop a numerical algorithm for solving equations in the LR models. It can also be considered as deriving "exact" wall-functions. We use this method for the model equation but it can be easily generalized to an arbitrary linear equation or system of equations under quite general assumptions. The main idea of the approach is as follows.

Near the wall in the domain $\Omega_1 = [0 \ \ y^*]$, the following two boundary-value problems is solved:

1. $Lu_1 = f \quad u_1(0) = u_0, \ du_1/dy\,(y^*) = 0 \qquad 0 \le y \le y^* \qquad (18)$
2. $Lu_2 = 0 \quad u_2(0) = 0, \ du_2/dy(y^*) = 1 \qquad 0 \le y \le y^* \qquad (19)$

In this case, $L \equiv \mu \dfrac{d^2}{dy^2} + y^n$, $f \equiv C$.

It is easy to prove that the general solution is

$$u(y) = u_1(y) + du/dy(y^*)u_2(y) \qquad (20)$$

If we consider (20) at point $y^*$, we have a Robin-type boundary condition for the rest domain $\Omega_2$: $y^* \le y \le 1$:

$$u(y^*) = u_1(y^*) + du/dy(y^*)\,u_2(y^*) \qquad (21)$$

This boundary condition is exact if we set it at $y = y^*$. If the "convective" term is neglected in the first problem, equality (21) exactly coincides with (14).

If we use the exact coefficient $\mu$ in the domain $\Omega_2$, we obtain the exact solution. In this case, we have some version of a decomposition method.

If we use the HR model in the domain $\Omega_2$ and define the boundary condition at $y^*$, the error is small, and the curves almost coincide. The difference is shown in Figure 6, using zoom:

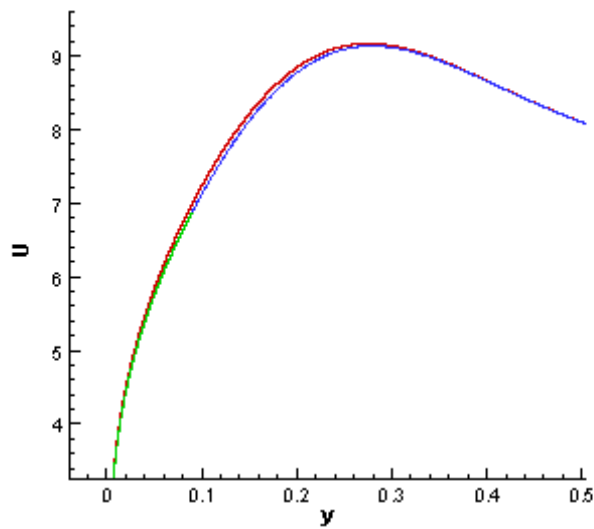

Fig. 6. Comparison of the exact (LR) solution and solution after decomposition.

In this case $y^* = 2\varepsilon$, $\tau_w = 0.18$. If the boundary condition to the wall is solved, we have: $\tau_w = 0.16$ if $y^* = \varepsilon$; $\tau_w = 0.19$ if $y^* = 2\varepsilon$, and $\tau_w = 0.21$ if $y^* = 3\varepsilon$.

The comparison of profiles is given in Figure 7:

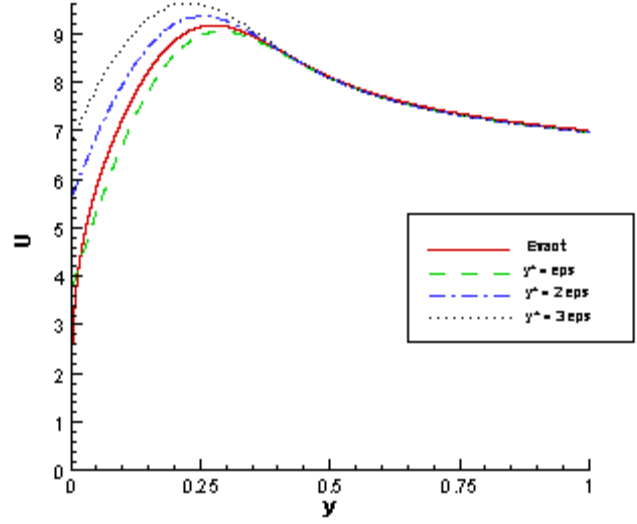

Fig. 7. Comparison with exact (LR) solution for different $y^*$.

If adopting this approach as a decomposition method to solve the LR equations, we have to solve two problems near the wall. On the other hand, we can easy optimize a mesh because the meshes in the domains $\Omega_1$ and $\Omega_2$ can be constructed completely independently.

*3.5.1. Newman problem*

In this case the algorithm is similar to the Dirichlet problem. We solve the following two boundary-value problems:

1.  $Lu_1 = f \quad du_1/dy(0) = u_0, \quad u_1(y^*) = 0$ (22)

2.  $Lu_2 = 0 \quad du_2/dy(0) = 0, \quad u_2(y^*) = 1$ (23)

The general solution is

$$u(y) = u_1(y) + u(y^*)u_2(y) \qquad (24)$$

After derivation, a Robin-type boundary condition at $y^*$ is obtained:

$$du/dy(y^*) = du_1/dy(y^*) + u(y^*)du_2/dy(y^*) \qquad (25)$$

We use this boundary condition in the domain $\Omega_2$ ($y^* \leq y \leq 1$), and in the domain $\Omega \backslash \Omega_2$ the solution is obtained from (24), upon $u(y^*)$ is known from the previous problem.

## 4. Conclusion

For a model equation simulating LR models, different wall-functions have been studied and derived. Our study has revealed that although the analytical wall-functions are less mesh-dependent than the standard wall-functions, some substantial mesh sensitivity does remain, especially on a fine mesh.

A new Robin-type interpretation of the wall-functions has been suggested. The boundary conditions (wall-functions) are mesh-independent in this case. There is some dependence on one or two free parameters (including the sub-layer thickness) but it is weak provided the parameters are reasonably estimated.

New analytical and numerical wall-functions of a general type have been derived. In the former case they are valid for any efficient "turbulence coefficient" $\mu$. In the latter case, the wall-functions can be treated as "exact" boundary conditions. The algorithm developed can be considered as a decomposition method and allows us to split the problem into a near-wall part and the rest one. Since the algorithm is exact, one can use it for effective solving the low-Reynolds-number equations directly.

Although the theory has been developed for a model equation, it can be used for solving the "real" LR equations including the major algorithms and formulas derived. In that case, the decomposition method is to be included into non-linear iterations. The integrals in (15) can be estimated from a previous iteration or time step. There is a room for optimization, e.g., the boundary-value problem (18) for the uniform equation can be solved once.

## 5. Acknowledgment

The author is grateful to A.V. Gerasimov for fruitful discussions, and D.R.Laurence for initiating this research and showing a constant interest in this work.

This work has been supported by the FLOMANIA project (Flow Physics Modelling - An Integrated Approach) is a collaboration between Alenia, AEA, Bombardier, Dassault, EADS-CASA, EADS-Military Aircraft, EDF, NUMECA, DLR, FOI, IMFT, ONERA, Chalmers University, Imperial College, TU Berlin, UMIST and St. Petersburg State Technical University. The project is funded by the European Union and administrated by the CEC, Research Directorate-General, Growth Programme, under Contract No. G4RD-CT2001-00613.